\documentclass[12pt]{article}
\usepackage{graphicx,color}
\usepackage[utf8]{inputenc}
\usepackage{a4,amssymb,amsmath}

\def\sig{\sigma} \def\kap{\kappa} \def\lam{\lambda}
\def\inv{^{-1}}
  \def\lra{\leftrightarrow}

\def \RR {{\mathbb R}}

\def \merw#1#2{{}_{#1}M^{#2}}
\def\AP{A^{\rm P}} \def\AK{A^{\rm F}}  
\def\tpf{2-point function}
\def\set{stress-energy tensor}
\def\wick#1{{:}#1{:}} 

\def\be{\begin{equation}} \def\bea{\begin{eqnarray}}
\def\ba{\begin{array}} \def\eea{\end{eqnarray}} 
\def\ee{\end{equation}}\def\ea{\end{array}}

 \numberwithin{allcount}{section}
\newcommand{\eref}[1]{Eq.~(\ref{#1})}
\newcommand{\sref}[1]{Sect.~\ref{#1}}

\begin{document}

\title{\sc Relations between Positivity, Localization and Degrees of Freedom: \\[2mm]
\large\sc the Weinberg-Witten theorem and the van~Dam-Veltman-Zakharov discontinuity}

\author{Jens Mund \\ {\small
Departamento de F\'isica, Universidade Federal de Juiz de Fora,} \\
{\small Juiz
 de Fora 36036-900, MG, Brasil, email: mund@fisica.ufjf.br} \\[3mm]
Karl-Henning Rehren \\ {\small Institut f\"ur Theoretische Physik,
  Universit\"at G\"ottingen,} \\ {\small 37077 G\"ottingen,
 Germany,  email: rehren@theorie.physik.uni-goettingen.de} \\[3mm] Bert
Schroer \\ {\small Centro Brasileiro de Pesquisas F\'isicas, 22290-180 Rio de Janeiro, RJ, Brasil;} \\ {\small Institut f\"ur Theoretische Physik der FU
  Berlin, 14195 Berlin, Germany,} \\ {\small email: schroer@zedat.fu-berlin.de} }

\maketitle

\begin{abstract}

The problem of accounting for the quantum degrees of freedom in
passing from massive higher-spin potentials to massless ones, and
the inverse problem of ``fattening'' massless
tensor potentials of helicity 
$\pm h$ to their massive $s=\left\vert h\right\vert $ counterparts,
are solved -- in a perfectly ghost-free approach -- using
``string-localized fields''. 

This approach allows to overcome the Weinberg-Witten impediment against the
existence of massless $\left\vert h\right\vert \geq 2$ energy-momentum
tensors, and to qualitatively and quantitatively resolve the van Dam-Veltman-Zakharov discontinuity concerning,
e.g., very light gravitons, in the limit $m\to0$. 
\end{abstract}

%\tableofcontents

\section{Introduction}
\label{s:intro}
In relativistic quantum field theory, the quantization of interacting
massless or massive classical potentials of
higher spin ($s\geq 1$) either violates Hilbert space positivity which is an
indispensable attribute of the probability interpretation of quantum theory,
or leads to a violation of the power counting bound of renormalizability
whose maintenance requires again a violation of positivity. 

In order to save positivity for those quantum fields which
correspond to classically gauge invariant observables, one usually formally
extends the theory by adding degrees of freedom in the form of negative
metric St\"{u}ckelberg fields and ``ghosts'' without a counterpart
in classical gauge theories. The justification for this quantum gauge
setting is that one can extract from the indefinite metric
Krein space a Hilbert space that the gauge invariant operators
generate from the vacuum.  

This situation is satisfactory as far as the vacuum sector and the
perturbative construction of a unitary gauge-invariant S-matrix are
concerned. However the theory remains incomplete in that it provides
no physical interpolating fields that mediate between the causal
localization of the field theory and the analytic structure of the
S-matrix in terms of fields that connect charged states with the vacuum. 
Expressed differently, gauge theory allows to compute the perturbative
S-matrix, but cannot construct its off-shell extension on a Hilbert space.

There are two famous results about the higher-spin massless case. The
first is the Weinberg-Witten theorem \cite{WW} which states that for
$s\geq2$, no point-localized \set\ exists such that the
Poincar\'e generators are moments of its zero-components.
This result also obstructs the semiclassical coupling of massless
higher-spin matter to gravity. 

The second is the DVZ observation due to van Dam and Veltman \cite{vDV} and to
Zakharov \cite{Z}, that in interacting models with $s\geq 2$,
scattering amplitudes are discontinuous in the mass at $m=0$, i.e.,
the scattering of matter through exchange of massless gravitons (say)
is significantly different from the scattering via gravitons of a very
small mass.  

Both problems can be addressed, without being plagued by the
positivity troubles   
of gauge theories, with the help of ``string-localized quantum
fields'' defined in the physical Hilbert space. The latter may be
regarded as a fresh start to Mandelstam's attempts \cite{Ma} to
reformulate gauge theories as full-fledged 
theories in which all fields live on the Hilbert space of
the field strength. The new point of view was triggered by a new
approach to Wigner's infinite-spin representation \cite{MSY}, that
proved to be useful also for finite spin. String-localized potentials
for finite spin $s$ are integrals over their field strengths along a
``string'' $x+\RR^+_0e$, see \eref{integral}, \eref{AFe}, or
\eref{A2-def}, where $e$ is a (spacelike) direction. This evidently
does not change the particle content. The main benefit
of these potentials is their 
improved UV dimension $d_{\rm UV}=1$ rather than $d_{\rm UV}=s+1$,  
admitting renormalizable interactions that are otherwise excluded by
power-counting.  

A point-localized massive spin $s$ potential can be split up
into a string-localized potential that has a massless limit, and
derivatives of one or more so-called ``escort fields''. The
role of the latter is to separate off derivative terms from the interaction
Lagrangean or from conserved currents, that do not contribute to
the S-matrix or to charges and Poincar\'e generators, 
respectively. They thus ``carry away'' all non-renormalizable UV
fluctuations and singularities in the limit $m\to0$.  

How this works, may be illustrated in the case of QED
\cite{S15,S16,M}: The coupling to the indefinite Maxwell potential
$\AK$ (``F'' stands for ``Feynman gauge'') is replaced by a coupling 
$j^\mu\AP_\mu$ to the massive Proca potential $\AP$. This avoids
negative-norm states, but the interaction is non-renormalizable because of 
the UV dimension 2 of the Proca potential. Now, the decomposition (see 
\sref{s:sl1}) $\AP_\mu(x)=A_\mu(x,e) - m\inv\, \partial_\mu a(x,e)$ into
a string-localized potential and its escort is
  brought to bear: $A_\mu(x,e)$ 
has UV dimension 1 and is regular at $m=0$. The UV-divergent part of
the interaction is carried away by the escort field:  
$-m\inv \, j^\mu\partial_\mu a(e) = -\partial_\mu (m\inv \,
j^\mu a(e))$ is a total derivative and may be discarded from the
interaction Lagrangean. The remaining string-localized (but equivalent
to the point-localized) interaction $j^\mu A_\mu(e)$ has UV dimension
4, and remains renormalizable at $m=0$.  

The ongoing analysis of perturbation theory with string-localized
interactions \cite{GMV,M,MS} gives strong evidence that the resulting
theory is order-by-order renormalizable, and equivalent to the
``usual'' QED. The scattering matrix can be made independent of the
string direction $e$, provided a suitable renormalization condition is
satisfied. In that case, interacting observable fields are
string-independent and hence local. These conditions can be seen as
an analogue of Ward identities imposed in order
  to ensure BRST invariance in point-localized but indefinite 
approaches \cite{Sf,DGSV}, see also footnote 1 in \sref{s:outlook}.
Indeed, the conditions can also be formulated in a cohomological
manner. Yet, the precise relation between gauge invariance and
string-independence remains to be explored.    

We give more details, especially on the preservation of
  causality, for the (much easier) case of the coupling 
  of a massive vector field to an external source in 
  \sref{s:caus}.

Whereas string-localized perturbation theory is still in its infancy,
the problems of massless currents and energy-momentum tensors as well
as the continuous passage from free massive fields to
their massless helicity counterparts can be completely solved. The presentation
of this solution is the principal aim of this letter, including also
the opposite direction, sometimes (in connection with the Higgs mechanism)
referred to as ``fattening''.

\subsection{Overview of results}
\label{s:overview}

We outline the general picture for arbitrary integer spin $s$, 
referring to \cite{MRS} for further details. As the case $s=2$ 
already exhibits all the features of the general case, we focus 
on $s=1$ and $s=2$ in \sref{s:sl1} and \sref{s:sl2}.

The \tpf s of covariant massless potentials are indefinite polynomials
in the metric tensor $\eta_{\mu\nu}$, while their field strengths 
(curl in all indices) are positive. (By ``positive'', it is understood
''positive-semidefinite'', accounting for null states due to equations
of motion like $\partial^\mu F_{\mu\nu}=0$.) 
Alternatively, the field strengths can be 
constructed, without reference to a potential, directly on the Fock
space over the unitary massless helicity $h=\pm s$ Wigner
representations of the Poincar\'e group.  
This is exposed in standard textbooks, e.g., \cite{W}. One can construct
potentials in the Coulomb gauge on the same Hilbert space, but one
gets into conflict with Poincar\'e covariance: Lorentz transformations
result in an operator-valued gauge transformation due to
the affine nature of the Wigner phase. When the potentials are
required for interactions, and one has to compromise
between positivity or Lorentz invariance, preference is usually
given to covariance.  

For some early treatments of massive free tensor fields of
  higher spin, see \cite{Fz,Fd}. We freely adopt the name ``Proca''
  for all spins $s\geq 1$. 
The Proca potentials are symmetric traceless and conserved tensors
$\AP_{\mu_1\dots\mu_s}(x)$ of rank $s$.
Their \tpf s obtained from the $(m,s)$ Wigner representation \cite{W} are
polynomials in the positive projection orthogonal to the momentum (sign convention $\eta_{00}=+1$) 
\bea\notag
-\pi_{\mu\nu}(p) = -\eta_{\mu\nu} + 
\frac{p_\mu p_\nu}{m^2}
\eea 
with coefficients dictated by
symmetry and tracelessness. The momenta in the numerator cause the UV
dimension $d_{\rm UV}=s+1$ and, by power counting, jeopardize the
renormalizability of minimal couplings to currents.  

The potentials evidently admit no massless limit. Only their
field strengths $F_{[\mu_1\nu_1]\dots[\mu_s\nu_s]}$ exist at $m=0$ because the curls kill the terms with momentum factors.

We define symmetric free tensor fields $A^{(r)}_{\mu_1\dots\mu_r}(x,e)$ of rank
$0\leq r\leq s$ on the Fock space of the massive field strengths such that 

\medskip

$\bullet$ All $A^{(r)}$ have UV dimension $d_{\rm UV}=1$ and are
  regular in the massless limit.

$\bullet$ The potential $\AP$ can be decomposed in a way that (i) all
contributions of UV dimension $>1$ are isolated as derivatives of the
``escort'' fields $A^{(r)}$ of lower rank $r < s$, and (ii) the
singular behaviour at $m\to0$ is manifest in the expansion coefficients
  (inverse powers of $m$).     

$\bullet$ The massive fields $A^{(r)}$ are
  coupled among each other through their traces and divergences. In
  the massles limit, they become traceless and conserved, 
  and their field equations and \tpf s decouple.

$\bullet$ At $m=0$, the escort $A^{(0)}$ is the canonical massless
  scalar $\varphi$. The tensors $A^{(r>0)}$ are potentials for the
  field strengths of helicity $h=\pm r$ \cite{W}. They were previously
  constructed \cite{PY} without an approximation from $m>0$.  

$\bullet$ Conversely, the given massless potential $A^{(s)}$ 
  of any helicity $h=\pm s$ can be made massive (``fattening'') in the
  same way as for scalar and Dirac fields, namely by
  simply changing the dispersion relation $p^0=\omega_m(\vec p)$. The
  fattened field brings along with it all lower rank fields $A^{(r)}$
  by virtue of the coupling through the divergence. We give a
  surprisingly simple formula involving only derivatives, to restore
  the exact Proca potential $\AP$ from the fattened $A^{(s)}$.

$\bullet$ The massless limit shows the way to construct a \set\ for the
massless fields that decouples into a direct sum of mutually commuting 
  \set s $T^{(r)}$ for the helicity potentials $A^{(r)}$.

$\bullet$ None of these constructions refers to a classical action
principle. The quantization is manifest and without ghosts from the outset. 

\medskip

The massless limit describes the exact splitting of the $(m,s)$ Wigner
representation into massless helicity representations with $h=\pm r$
($r=1,\dots,s$) and $h=0$. 

In particular, the number $2s+1$ of one-particle states at fixed
momentum is preserved. In contrast, the ``fattening'' of
the massless potential of helicity $s$ increases the number of one-particle
states, because its \tpf\ is a semi-definite quadratic form of rank
$2$ that becomes rank $2s+1$ under the deformation of the dispersion relation.

These facts yield the obvious explanation of the DVZ discontinuity
\cite{vDV,Z} in linearized gravity coupled to external
sources: The spin 2 Proca potential $\AP$ (or its analog in the 
indefinite Feynman gauge) is not continuously connected with a
massless helicity $h=\pm 2$ potential. At each positive mass, the 
former has contributions from all $r\leq 2$. Rejecting at $m=0$ the
helicities $\vert h\vert < 2$ causes the discontinuity. We shall
exhibit in \sref{s:sl2} that in the ghost-free setting, at $m=0$ only
the helicities $h=\pm2$ (the linearized massless gravity) and $h=0$
survive in the coupling to the external source. The $h=\pm2$ part is
the linearized massless gravity. The additional scalar $\varphi(x)$ is
the massless limit of the scalar escort field $A^{(0)}(x,e)$ and
couples to the trace of the stress-energy tensor.

These results state the preservation of degrees of freedom (of free
fields coupled to external sources) in a ghost-free language. The DVZ
discontinuity 
arises by dropping the scalar ``by hand'' at $m=0$. In contrast, there
exist several ideas to explain how the scalar, and hence the
discontinuity, could be instead dynamically suppressed in the presence
of suitable interactions. Notably, Vainshtein \cite{Vs} has presented a
model with a non-perturbative screening mechanism that is effective
only at small distances. It would be extremely rewarding to see how
this screening emerges in a ghost-free approach. Other authors appeal
to curvature effects or extra dimensions, c.f.\ the review
\cite{H}. We hope that our physical approach to free fields will also
contribute to a better understanding of the interacting models.

The stated properties of the massless potentials and
\set s are clearly at variance with many No-Go theorems, including the
Weinberg-Witten theorem. This is possible because they are
string-localized. 
Their \tpf s involve, instead of the singular (as $m\to0$) tensor
$\pi_{\mu\nu}(p)$ or indefinite tensor $\eta_{\mu\nu}$, a suitable
tensor $E_{\mu\nu}(p)$ whose substitution into the \tpf s (i)
preserves positivity, (ii) does not affect the field strengths, and
(iii) has a regular limit $m\to0$.  

The No-Go theorems may be attributed to the fact that such a tensor
$E_{\mu\nu}(p)$ does not exist, if it is allowed to be a function of
the momentum only. Instead,  
\bea\notag
E(e,e')_{\mu\nu}(p):= 
\eta_{\mu\nu} - \frac{p_\mu
  e_\nu}{(pe)_+}-\frac{e'_\mu p_\nu}{(pe')_+} + \frac{(ee')p_\mu
  p_\nu}{(pe)_+(pe')_+} 
\eea
(where $i/(k)_+=i/(k+i0)$ is the
Fourier transform of the Heaviside function) are distributions in $p$ and two 
four-vectors $e$, $e'$. If $E_{\mu\nu}$ is substituted
for $\pi_{\mu\nu}$ or $\eta_{\mu\nu}$, the potentials depend on
$e$, but the field strengths will not.

In momentum space, the integration  
\bea\label{integral}
X(x,e) \equiv (I_e X)(x) := \int_0^\infty d\lam\, X(x+\lam e)
\eea
produces the denominators $i((pe)+i0)\inv$ in the
creation part and $-i((pe)-i0)\inv$ in the annihilation
part. Thus, fields whose \tpf s are polynomials in $E_{\mu\nu}$
are necessarily localized along the ``string'' $x+\RR^+_0e$.

String-localization requires some comments. First, it is not a
feature of the associated particles, but of the fields that may be
used to couple them to other particles. (The only exception are
particles in the infinite-spin representations \cite{MSY,LMR}, that
are beyond the scope of this letter.) 

\eref{integral} (and its generalizations involving  several integral 
operations $I_e$) imply the Poincar\'e transformations of
string-localized fields   
\bea\notag\label{covariance}
U_{a,\Lambda} A_{\mu_1\dots\mu_r}(x,e) U_{a,\Lambda}^* =
\big(\prod\nolimits_i\Lambda^{\nu_i}{}_{\mu_i}\big)A_{\nu_1\dots\nu_r}(a+\Lambda x,\Lambda
e),\quad
\eea
i.e., the direction of the string is transformed along with its apex
$x$ and the tensor components of the field tensor. 

There is no conflict with the principle of causality, which is as
imperative in relativistic quantum field theory as Hilbert space positivity. 
String-localized fields satisfy causal commutation relations according
to their localization: two fields commute whenever their strings
are pointwise spacelike separated. There are
sufficiently many spacelike separated pairs of spacelike or lightlike
strings to construct scattering states by asymptotic cluster
properties (Haag-Ruelle theory). 
%%CHANGE
For this reason, scattering theory
requires $e^2\leq 0$.

String-localized interactions admit couplings of physically massive
tensor potentials without spontaneous symmetry breaking (cf.\
\sref{s:outlook}). Instead, when coupling self-interacting massive
vector bosons (like $W$ and $Z$ bosons) via their string-localized
potentials, the string-independence can only be achieved with the help
of a boson with properties like the Higgs, including a quartic
self-interaction \cite{S16}. Its role is, however, not the generation
of the mass, but the preservation of the renormalizability and
locality.

Examples of new renormalizable interactions in the string-localized
setting could be the coupling of matter to gravitons through
the string-localized potentials $A^{(2)}$, and perhaps the
self-coupling of gravitons.

In the sequel, we give more details for spin 1 and 2.
All displayed linear relations between fields follow from their
definitions by integrals and derivatives of point-localized fields,
e.g., by inspection of their integral representations in terms of
creation and annihilation operators. 

We write \tpf s throughout as
$$(\Omega,X(x)Y(y)\Omega)= \int d\mu_m(p) \cdot 
e^{-ip(x-y)}\cdot \merw{m}{X,Y}(p),$$
where $d\mu_m(p)=\frac{d^4p}{(2\pi)^3}\delta(p^2-m^2)\theta(p^0)$.

\section{Spin one}
\label{s:sl1}
\setcounter{equation}{0}

The \tpf\ of the massless Feynman gauge potential 
\bea
\merw{0}{\AK_\mu,\AK_\nu}= -\eta_{\mu\nu}
\notag
\eea 
is indefinite. Its curl 
$F_{\mu\nu}=\partial_\mu A_\nu-\partial_\nu A_\mu$ is the Maxwell
field with positive \tpf\  
\bea
\merw{0}{F_{\mu\nu}, F_{\kap\lambda}} =
-p_\mu p_\kap\,\eta_{\nu\lambda} + p_\mu p_\lam\,\eta_{\nu\kap}+p_\nu p_\kap\,\eta_{\mu\lambda} - p_\nu p_\lam\,\eta_{\mu\kap}.
\notag
\eea

The massive Proca potential satisfies $\partial^\mu\AP_\mu=0$. Its
positive \tpf\ is  
\bea
\merw{m}{\AP_\mu, \AP_\nu}= -\pi_{\mu\nu}(p).
\eea
The curl kills the term $p_\mu p_\nu/m^2$, so the field strength is
regular at $m=0$.  
The field equation $\partial^\mu F_{\mu\nu} = -m^2 \AP_\nu$
gives back the potential in terms of its field strength.

Only for $s=1$, the massless limit can be achieved with point-localized fields:
by inspection of their \tpf s, $m\AP_\mu$ is regular at $m=0$, where it
decouples from $F_{\mu\nu}$ and becomes the derivative of the
canonical scalar
free field $\varphi$ with
$\merw{0}{\partial_\mu\varphi,\partial_\nu\varphi} = p_\mu p_\nu$.  

In the string-localized setting, the massless scalar emerges without
derivative. We define
\bea\label{AFe} 
A^{(1)}_\mu(x,e)\equiv A_\mu(x,e)&:=& \big(I_eF_{\mu\nu}\big)(x)e^\nu\equiv\int\nolimits_0^\infty 
d\lam\, F_{\mu\nu}(x+\lam e)e^\nu,\\ 
A^{(0)}(x,e)\equiv a(x,e)&:=&-m\inv\, \partial^\mu A_\mu(x,e).
\eea
$A_\mu(e)$ is regular in the massless limit because $F_{\mu\nu}$ is. 
That
$a(e)$ is also regular can be seen from 
\bea\label{AA1}
\merw{m}{A_\mu(-e), A_\nu(e')} = - E(e,e')_{\mu\nu}(p), \quad
\eea
which implies by the definition of $a(e)$
\bea\notag
\merw{m}{a(-e), A_\nu(e')} = O(m), \qquad
\merw{m}{a(-e),a(e')} = 1 +O(m^2).
\eea
(As the fields are distributions also in
$e$ \cite{MO}, we have to admit independent string directions $e$,
$e'$. The choice ``$-e$'' is a convenience paying off for higher spin \cite{MRS}.)

At $m=0$, the fields $a$ and $A_\mu$ decouple, and converge to the
massless scalar and (as the terms $O(p/(pe))$ in \eref{AA1} 
do not contribute to $F_{\mu\nu}$) to a string-localized massless
potential for the Maxwell field strength. 

In addition, one gets the decomposition underlying the QED 
example in \sref{s:intro} 
\bea\label{expans1}
\AP_\mu(x)= A_\mu(x,e) - m\inv\, \partial_\mu a(x,e).
\eea
The taming of the UV behaviour is seen from \eref{AA1}:
the momentum factors in the denominators of $E(e,e')$ balance those in
the numerators \cite{MO}.

The string-localized field $A_\mu(x,e)$ for
$e=(1,\vec 0)$ coincides with the Coulomb gauge field $A^{\rm C}_\mu$.
The well-known non-locality of the Coulomb gauge potential reflects
the fact that two timelike strings are never spacelike separated. Its
failure of covariance (when $e_0$ is fixed) is due to \eref{covariance} 
which requires an additional gauge transformation to bring $\Lambda
e_0$ back to $e_0$.
  
It may also be interesting to notice that one can average the 
potential $A_\mu(x,e)$ in $e$ over the spacelike sphere with $e^0=0$. The
resulting field is again the Coulomb gauge potential.

Similarly, for fixed spacelike $e$, $A_\mu(e)$ coincides with an axial
gauge potential satisfying $e^\mu A_\mu=0$. However,
this relation is not used as a gauge condition to
reduce the degrees of freedom before quantization, but instead 
the potentials for all $e$ coexist simultaneously on the Fock space of
the field strength, and they covariantly transform into each other
according to \eref{covariance}.  
By specifying the \tpf\ for spacelike $e$ as a distribution rather
than a function with a singularity, the manifestly
string-localized representation \eref{AFe} of the axial gauges is revealed, and
the mutual commutativity of axial gauge fields for
different directions is discovered. 

\subsection{Preservation of causality}
\label{s:caus}
One might worry that a string-localized interaction Lagrangean could spoil the
causality of the resulting perturbation theory. We sketch here why
this does not happen. We choose the easiest and most transparent
example: the interaction of a massive vector field with a conserved
external (classical) current $j^\mu(x)$. It is essential that the
string-dependence of the interaction Lagrangean is a total derivative: 
\bea\label{LV} L_{\rm int}(x,e) = A_\mu(x,e)j^\mu(x) = \AP_\mu(x)j^\mu(x)
+\partial_\mu \big(\phi(x,e)\,j^\mu(x)\big)
\eea
so that the classical action, and hence the lowest order of the
S-matrix, is independent of $e$.
(In the massless case, neither $\AP$ nor $\phi(x,e)=m\inv a(x,e)$
exist, but the variation of $L_{\rm int}(e)$ w.r.t.\ $e^\kappa$ is still a
total derivative, because $\partial_{e^\kappa}A_\mu(x,e)=\partial_\mu
(I_eA_\kappa)(x)$, where $I_e$ is the integral along the string as in \eref{AFe}. This crucial feature is shared by many other interactions of
interest \cite{S15,S16,GMV}, cf.\ examples in \sref{s:outlook}, where
the renormalizability conditions in higher orders are less trivial
than in the present external source problem.)

In order that the causal S-matrix 
$$S_e[gj]:= T\exp \, i \int d^4x \, g(x)\, A_\mu(x,e)j^\mu(x)$$
is independent of $e$ in the limit $g\to const$, the decomposition
\eref{LV} must continue to hold ``under the time-ordering''. 
This can be formulated as a condition on the string-localized Feynman
propagator $i(\Omega,TA_\mu(x,e)A_{\mu'}(x',e')\Omega)$, which is quite
non-trivial because the time-ordering must 
be taken along the strings $x+\RR^+_0e$, $x'+\RR^+_0e'$, and the
singularities at intersections of strings have to be carefully
analyzed and renormalized \cite{MO,M}. The condition can be fulfilled and gives
the unique answer
$$ i(\Omega,TA_\mu(x,e)A_{\mu'}(x',e')\Omega) = (-\eta_{\mu\mu'}
- \partial_\mu e_{\mu'} I_{e} + \partial_{\mu'} e'_\mu I_{-e'} +
(ee')\partial_\mu\partial_{\mu'} I_{e}I_{-e'}) G_F(x-x'),$$
where $G_F(x-x')$ is the scalar Feynman propagator.

We proceed by computing the interacting potential in the setting of
causal perturbation theory based on Bogoliubov's formula \cite{Bg}
$$
A^{\rm int}_\mu(x,e) := S_{e_0}[j]\inv\frac{-i\delta}{\delta f^\mu(x,e)}
S_{e_0}[j,f]\Big\vert_{f=0}
$$
where $S_{e_0}[j,f] = T \exp \, i\int d^4x \big(A_\mu(x,e_0)j^\mu(x) +
\int d\sigma(e) A_\mu(x,e)f^\mu(x,e)\big)$. In this approach,
renormalization amounts to the proper definition (as distributions) of
propagators and their products. As it can be done in position space, it
is best suited to control causality. In the external source problem,
no further renormalization is necessary. Namely, by Wick's Theorem,
we get
$$A^{\rm int}_\mu(x,e) = A_\mu(x,e) + \int d^4x' \,g(x)\,G^{\rm
  ret}_{\mu,\mu'}(x,e;x',e_0) j^{\mu'}(x')$$
where the string-localized retarded Green function $G^{\rm ret} = 
i(\Omega,(T[AA']- A'A)\Omega)$ equals 
$$G^{\rm ret}_{\mu,\mu'}(x,e;x',e_0) = (-\eta_{\mu\mu'}
- \partial_\mu e_{\mu'} I_{e} + \partial_{\mu'} e_{0\mu} I_{-e_0} +
(ee_0)\partial_\mu\partial_{\mu'} I_{e}I_{-e_0}) G^{\rm ret}(x-x').$$
The contributions depending on $e_0$ vanish in the limit $g\to const$,
because $j^\mu$ is conserved, hence $A^{\rm int}_\mu(x,e)$ is
independent of $e_0$. The remaining contributions can be written as 
$$A^{\rm int}_\mu(x,e) = A_\mu(x,e) + \big(A^{\rm cl}_\mu(x)
+ \partial_\mu \phi^{\rm cl}(x,e)\big)\cdot \mathbf{1}$$
where 
$$A^{\rm cl}_\mu(x)=-g\int d^4x \, G^{\rm ret}(x-x') j_\mu(x'),\qquad \phi^{\rm cl}(x,e)=-g\int d^4x \, G^{\rm ret}(x-x') j(x',e) $$
are classical fields with sources $j^\mu(x)$ and $j(x,e)=e_\mu
\int_0^\infty d\lambda j^\mu(x+\lambda e)$, respectively. The field
strength is then manifestly independent of $e$, and coincides with the
solution in the point-localized setting -- except that the latter has a
$\delta$-function ambiguity for the Feynman propagator of the Proca
field due to its bad $UV$ behaviour. In the string-localized setting,
the ambiguity is fixed ($=0$). For more details, and for the QED case
with a quantum source, see \cite{M}.

\section{Spin two}
\label{s:sl2}
\setcounter{equation}{0}

The case $s=2$ is largely analogous, but the decoupling at $m=0$
requires a second step. 

The positive \tpf\ of the massless field strength
$F_{[\mu\kap][\nu\lam]}$ can be represented as the curl of the
(auxiliary) indefinite \tpf\ of the Feynman gauge potential 
\bea\label{AA2-K}
\merw{0}{\AK_{\mu\nu}, \AK_{\kap\lam}}
=\frac12\big[\eta_{\mu\kap}\eta_{\nu\lam}+\eta_{\mu\nu}\eta_{\kap\lam}\big]
-\frac12\eta_{\mu\nu}\eta_{\kap\lam}.\qquad
\eea 
The coefficient $-\frac12$ of the last term ensures that
there are precisely two helicity states. 

The symmetric, traceless and conserved massive Proca \tpf\ is 
\bea\label{AA2-m}
\merw{m}{\AP_{\mu\nu},\AP_{\kap\lam}}
=\frac12\big[\pi_{\mu\kap}\pi_{\nu\lam}+\pi_{\mu\lam}\pi_{\kap\nu}\big]
-\frac13\pi_{\mu\nu}\pi_{\kap\lam}.\qquad
\eea 
The coefficient $-\frac13$ of the last term ensures the vanishing of
the trace. The formulae for the massive and massless field strengths
differ {\em only}
by this coefficient. In particular, the massless field
strength is not the limit of the massive field strength as $m\to0$.

In the string-localized setting, we define the massive potential
\bea\label{A2-def}A_{\mu\nu}(x,e):=\big(I_e^2
F_{[\mu\kap][\nu\lam]}\big)(x)e^{\kap} e^{\lam}
\eea
with $I_e$ as in \eref{AFe} iterated twice, and its escort fields 
\bea\notag\label{a2-def}
a^{(1)}_\mu(x,e)&:=&-m\inv\,\partial^\nu 
A_{\mu\nu}(x,e), \\ a^{(0)}(x,e)&:=& -m\inv\,\partial^\mu a^{(1)}_\mu(x,e).
\eea
\eref{AA2-m} implies
\bea\label{AA2-sl} 
\merw{m}{A_{\mu\nu}(-e),A_{\kap\lam}(e')} =
\frac12\big[E(e,e')_{\mu\kap}E(e,e')_{\nu\lam}+(\kap\lra\lam)\big]  
- \frac13 E(e,e)_{\mu\nu}E(e',e')_{\kap\lam},\quad
\eea
and one obtains the escort correlations with \eref{a2-def}. The
correlations between even and odd rank fields are $O(m)$ and 
decouple in the massless limit. The odd-odd and even-even correlations become
\bea\notag\label{Aa20}
\merw{0}{a^{(1)}_{\mu}(-e),a^{(1)}_\nu(e')} &=&
- \frac12 E(e,e')_{\mu\nu}(p),  \\
\merw{0}{A_{\mu\nu}(-e),a^{(0)}(e')} &=&
- \frac13 
E(e,e)_{\mu\nu}(p),  \\\notag
\merw{0}{a^{(0)}(-e),a^{(0)}(e')} &=& \frac23
\eea 
up to $O(m^2)$. $A_{\mu\nu}(e)$ and $a(e)$ do not decouple
at $m=0$, in fact one has $\eta^{\mu\nu} A_{\mu\nu}(e) = -a(e)$. 
In order to decouple the fields, notice that the operator
$$E_{\mu\nu}(e,e) = \eta_{\mu\nu}
 + (e_\nu\partial_\mu+e_\mu\partial_\nu) I_e 
+e^2 \partial_\mu\partial_\nu I_e^2$$
acts in momentum space on the creation and annihilation parts by
multiplication with $E(e,e)_{\mu\nu}(p)$ and with $E(e,e)_{\mu\nu}(-p) =
E(-e,-e)_{\mu\nu}(p)$, respectively. Thus, 
\bea\label{A22-def}
A^{(2)}_{\mu\nu}(e) := A_{\mu\nu}(e) +\frac12
E_{\mu\nu}(e,e)\, a^{(0)}(e)\quad\eea
decouples from $a^{(0)}$, and its \tpf\ is 
the same as \eref{AA2-sl} but with the proper coefficient $-\frac12$ rather than
$-\frac13$ for the last term. Thus, taken at $m=0$, $A^{(2)}$ is a
string-localized potential for the massless field strength
to which it is related by the same formula as \eref{A2-def}. It is,
unlike other potentials, positive, traceless and conserved.

In the massless limit, $A^{(0)}(e)=\sqrt{3/2}\,a^{(0)}(e)$ becomes
the $e$-independent massless scalar field by \eref{Aa20}. 
$A^{(1)}_\mu := \sqrt2\,a^{(1)}_\mu$ is the same string-localized
Maxwell potential as obtained from $s=1$.

The generalization of \eref{expans1} 
\bea\label{AAP}
\AP_{\mu\nu}(x) &=& 
A^{(2)}_{\mu\nu}(x,e) -\sqrt{1/6}\, E_{\mu\nu}(e,e)\, A^{(0)}(x,e)
 - \\ \notag &&-\frac{\sqrt{1/2}}{m}\,\big(\partial_\mu A^{(1)}_\nu
+\partial_\nu A^{(1)}_\mu\big)(x,e) +
\frac{\sqrt{2/3}}{m^2}\, \partial_\mu\partial_\nu A^{(0)}(x,e).
\eea
quantifies the singular lower helicity contributions to $\AP$.
 
Now, turning to the DVZ problem, we may couple
linearized massive gravity in a Minkowski background to a conserved stress-energy source by  
\bea\label{Se}
S_{\rm int}(e) = \int d^4x\, A_{\mu\nu}(x,e)T^{\mu\nu}(x).
\eea
Because by \eref{AAP}, $A_{\mu\nu}(e)$ differs from $\AP_{\mu\nu}$ only
by derivatives, the action is independent of $e$. At $m>0$, all five
states of the graviton couple to the source. In the limit $m\to0$, we
have by \eref{A22-def} 
$$A_{\mu\nu}(x,e) = A^{(2)}_{\mu\nu}(x,e) -
\sqrt{1/6}\,\eta_{\mu\nu} \,\varphi(x) + \hbox{derivatives},$$
where $\varphi(x)=\sqrt{3/2}\,\lim_{m\to0} a^{(0)}(x,e)$ is the massless
scalar field decoupled from the helicity-2 potential $A^{(2)}(x,e)$. Thus, 
\bea\label{Se0}
\lim_{m\to0} S_{\rm int}(e) = \int d^4x\,
A^{(2)}_{\mu\nu}(x,e)T^{\mu\nu}(x) - \sqrt{1/6} \int d^4x\,
\varphi(x)\, T_\mu^\mu(x).
\eea
The first term satisfies the $L$-$Q$ condition (see \sref{s:outlook})
and therefore equals the $e$-independent action for
linearized pure massless gravity. With a classical source, 
it can be treated exactly as in \sref{s:caus}.

We have thus (along with the known decoupling of the helicity $\pm1$
degrees of freedom) explicitly identified the scalar field that is
responsible for the DVZ discontinuity, as the limit of the escort
field on the massive Hilbert space. Our result is formally similar 
to Zakharov's who writes (in an indefinite gauge) instead the massless coupling 
$\AK_{\mu\nu}(x,e)T^{\mu\nu}(x)$ as the limit of the massive coupling
plus a compensating scalar ghost \cite{Z}. We want to emphasize the
change of perspective when one avoids unphysical ghost degrees of freedom.

As at spin 1, also the spin-2 potentials $A_{\mu\nu}(e)$ and
$A^{(2)}_{\mu\nu}(e)$ at fixed $e$ can be regarded as axial gauges. 
The averaging over the string directions with $e^0=0$ is only possible 
for $A^{(2)}_{\mu\nu}(e)$ at $m=0$, and yields again the radiation
gauge potential $A^{\rm C}_{0\mu}(x)=0$.  

The case of general integer spin \cite{MRS} is very similar to $s=2$,
except for the more involved combinatorics. 

\section{String-localized \set}
\label{s:SET}
\setcounter{equation}{0}

The \set\ is by no means unique. It must be conserved and symmetric so
that the generators 
\bea\notag 
P_\sig = \int_{x_0=t}\! d^3x \, T_{0\sig}, \quad
M_{\sig\tau} = \int_{x_0=t} \! d^3x \, (x_\sig T_{0\tau} - x_\tau
T_{0\sig})
\notag
\eea
are independent of the time $t$; and the commutators with the
generators must implement the infinitesimal Poincar\'e transformations
given by the Wigner representation. (The commutators are fixed by
the \tpf s.) But one may add ``irrelevant''
local terms as long as they do not change the generators. 

One choice of a \set\ that produces the correct generators is the
``reduced \set'' ($\times=\mu_2\dots\mu_s$ is a multi-index)
\bea\label{redset}
T^{\rm red}_{\rho\sig}:= (-1)^{s}\Big[-\frac14
\wick{\AP_{\mu\times}\stackrel\lra{\partial_\rho}\stackrel\lra{\partial_\sig}
\AP{}^{\mu\times}} - \frac s2\, \partial^\mu\Big(\wick{\AP_{\rho\times}\stackrel{\lra}{\partial_\sig} \AP_\mu{}^{\times}}+(\rho\lra\sig)\Big)\Big].
\eea
It differs by ``irrelevant terms'' from the Hilbert
\set, defined as
the variation of a suitable generally covariant action w.r.t.\ the metric.  
The first term in \eref{redset} also appears in \cite{Fz}.
The second term does not contribute to the momenta, but
is needed to ensure the correct Lorentz transformations \cite{MRS}.

Expanding $\AP$ into $A^{(r)}(e)$ resp.\ $A^{(r)}(e')$, and
discarding irrelevant terms (involving derivatives of escort fields) that
``carry away'' all singularities when $m\to0$, one gets a
string-localized \set\ that admits a massless limit. Discarding more
terms that are irrelevant at $m=0$, one decouples it as the sum over
$r\leq s$ of 
\bea\label{Tr}
T^{(r)}_{\rho\sig}(e,e') = (-1)^{r}\Big[-\frac 14
\wick{A^{(r)}_{\mu\times}(e)\stackrel\lra{\partial_\rho}\stackrel\lra{\partial_\sig}
A^{(r)}{}^{\mu\times}(e')}- \frac
r4\,\partial^\mu\Big(\wick{A^{(r)}_{\rho\times}(e)\stackrel{\lra}{\partial_\sig}
  A^{(r)}_\mu{}^{\times}(e')}\ba{c}+(e\lra e')\\+(\rho\lra\sig)\ea\Big)\Big]\quad
\eea
understood as distributions in two independent directions $e,e'$. As in \eref{A2-def},
\bea\notag
A^{(r)}_{\mu_1\dots\mu_r}(x,e) = (I_e^r
F^{(r)}_{[\mu_1\nu_1]\dots[\mu_r\nu_r]})(x)\,e^{\nu_1}\dots
e^{\nu_r}
\eea 
can be expressed in terms of the massless field strengths.

As the massless potentials $A^{(r)}$ mutually commute, the generators
defined by $T^{(r)}$ separately implement the Poincar\'e transformations of $A^{(r)}$. 
Massless higher-spin currents of charged potentials are constructed similarly. For details
see \cite{MRS}.   

That the Weinberg-Witten theorem can be evaded with non-local
densities, was pointed out earlier in \cite{Lp}, where examples
with unpaired helicities were given. \eref{Tr} involving
string integrals over field strengths is perhaps the most conservative
alternative, also in comparison with other proposals to
couple massless higher-spin matter to gravity \cite{FV,V,BBS,H}.

\section{``Fattening''}
\label{s:fat}
\setcounter{equation}{0}

The \tpf s of the massless and massive string-localized potentials $A^{(s)}$ (for
any spin) are the same polynomials in the tensor
$E_{\mu\nu}(p)$, except that the argument $p$ of the functions $E_{\mu\nu}$ 
is taken on the respective mass-shell. Thus, one obtains the 
massive field $A^{(s)}$ from the massless field $A^{(s)}$ just by
changing the dispersion relation $p^0=\omega_m(\vec p)$. As the
massive \tpf\ was constructed on the Hilbert space of the Proca
potential, this deformation preserves positivity.   
Through the coupling to the lower escort fields, it
brings back all spin components of the Proca field. Indeed, the latter
is restored from the massive potential $A^{(s)}$ by
$$\AP_{\mu_1\dots\mu_s}(x) = (-1)^s\merw{m}
{\AP_{\mu_1\dots\mu_s},\AP{}^{\nu_1\dots\nu_s}} \, A^{(s)}_{\nu_1\dots\nu_s}\big\vert_m(x,e),$$
where in this formula $\merw{m}{\AP,\AP}$ is understood as a differential operator
(a polynomial in $\pi_{\mu\nu}=\eta_{\mu\nu} +m^{-2}\partial_\mu\partial_\nu$).

\section{Outlook: interactions}
\label{s:outlook}
\setcounter{equation}{0}

A crucial question is which physical interactions that otherwise
  are non-renormalizable or cannot be formulated on a Hilbert space,
  are accessible by couplings to string-localized fields. Although the
  actual perturbation theory is not the subject of this letter, we
  give an overview of possible interactions.

One class of interactions (called ``$L$-$V$-pairs'') are of the form
$$L_{\rm int}(e) = L_{\rm int} + \partial_\mu V^\mu(e)$$
where $L_{\rm int}$ is a possibly non-renormalizable string-independent
interaction Lagrangean, and the string-localized $L_{\rm int}(e)$ is
renormalizable. The string-dependent derivative term $\partial V$
disposes of the strong short-distance fluctuations of $L_{\rm int}$, typically by
means of escort fields. 

Further constraints may arise in order to secure the $e$-independence
of the perturbative S-matrix in higher orders in the 
coupling constant, and higher order interactions may be needed. We
refer to these as ``induced'' interactions.   

All couplings
$A_{\mu_1\dots\mu_s}(x,e)j^{\mu_1\dots\mu_s}(x)$ of massive potentials 
to conserved currents are of $L$-$V$ type, but also cubic
self-couplings of massive vector bosons  
$$f_{abc}\,F^a_{\mu\nu}(x)A^b{}^\mu(x,e) A^c{}^\nu(x,e) + m^2 f_{abc}\,
A^{a\nu}(x,e)\partial_\nu \phi^b(x,e)\phi^c(x,e)
$$
where $\phi^a(e)$ are the escort fields as in \sref{s:caus}, and
$f_{abc}$ is totally anti-symmetric; or more general expressions
admitting vector bosons ($W$ and $Z$) of different masses. 
A second order constraint is that $f_{abc}$ must satisfy the Jacobi
identity, so they are the structure constants of some Lie algebra
(without a gauge principle having been imposed). Induced
interactions in this case are the quartic Yang-Mills terms, as well as
an additional coupling to a Higgs boson of arbitrary mass and with a 
potential such that $\frac{m_H^2}2H^2+V(H)$ is the usual Higgs
potential with one of its minima at $H=0$\footnote{Some of these
  claims have not yet been conclusively 
  established, but work is in progress \cite{MS}. They are true in the
  analogous BRST approach pursued by Scharf et al.\ \cite{Sf,DGSV}, and up
  to now the same patterns are always repeated in the string-localized
  approach, where $e$-independence replaces BRST invariance.}.  

While in the non-abelian case the Higgs coupling is induced from the
cubic self-coupling of the vector bosons, an abelian Higgs
coupling $A_\mu(x,e)A^\mu(x,e) H(x)$ may be chosen directly provided
it is completed to a power-counting renormalizable $L$-$V$-pair 
$$A_\mu(e)A^\mu(e) H +
A_\mu(e)[\phi(e)\stackrel{\lra}{\partial^\mu}H]-\frac{m_H^2}2\phi(e)^2H
= \AP_\mu\AP{}^\mu H + \partial_\mu V^\mu$$
with $V^\mu = A^\mu(e)\phi(e)H+\frac12
\phi(e)^2\stackrel{\lra}{\partial^\mu} H$.
A Higgs potential is induced in higher orders.

A more general class (called ``$L$-$Q$-pairs'') are interaction
Lagrangeans $L_{\rm int}(e)$, for which 
$$\partial_{e^\kappa} L_{\rm int}(e) = \partial_\mu Q_\kappa^\mu(e)$$
holds, which is sufficient to secure the $e$-independence of $S_{\rm
int}(e)= \int d^4x\, L_{\rm int}(x,e)$. These include couplings
$A_{\mu_1\dots\mu_s}(x,e)j^{\mu_1\dots\mu_s}(x)$ of massless 
potentials to conserved (classical or quantum) sources, for which a
point-localized potential on the Hilbert space does not exist. Namely,
$\partial_{e^\kappa} A_{\mu_1\dots\mu_s}(x,e)$ is a sum of gradients. A
general understanding of 
$L$-$Q$ pairs, and which terms they induce in higher orders in order to
maintain renormalizability, is presently under investigation.

Perturbative correlation functions of observables (composite fields that are $e$-independent in the free theory) remain $e$-independent when the
interaction is switched on. This has to be secured by Ward
identities, that, e.g., allow to pull derivatives out of time-ordered
products. 

\section{Conclusion}
\setcounter{equation}{0}

We have identified string-localized potentials for massive particles of
integer spin $s$ on the Hilbert space of their field strengths, that
admit a smooth massless limit to decoupled potentials with helicities
$h=\pm r$, $r\leq s$. We have presented an inverse ``fattening''
prescription via a manifestly positive deformation of the \tpf. The
approach provides a way around the Weinberg-Witten theorem, and
explicitly and quantitatively exhibits the origin of the DVZ
discontinuity.  

Our results also allow to approximate string-localized fields in
  the massless infinite-spin Wigner representations \cite{MSY} by the
  massive scalar escort fields $A^{(0)}$ of spin $s\to\infty$,
  $m^2s(s+1)=\kappa^2=$ const.\ (Work in progress \cite{R-PL}.)

String-localized fields are a device to formulate quantum interactions
in terms of a given particle content, that allow to take into full
account the well-known conflicts between point-localization and
positivity. With their use, positivity is manifest, while
localization is controlled by renormalized causal perturbation
theory, as presently investigated in \cite{GMV,MS,M}. It bears formal 
analogies with BRST renormalization, but is more economic (avoiding
unphysical degrees of freedom), and much closer to the fundamental
principles of relativistic quantum field theory.   

It was shown in the framework of algebraic quantum field theory, that
to connect scattering states with the vacuum, may in certain theories
require operations localized in narrow spacelike cones; and in the
presence of a mass gap it cannot be worse than that \cite{BF}. The
emerging perturbation theory using string-localized fields is the practical
realization of this insight.

\bigskip

{\bf Acknowledgments:}
JM and KHR were partially supported by CNPq. KHR and BS 
enjoyed the hospitality of the UF de Juiz de Fora. We thank
D. Buchholz for pointing out ref.\ \cite{Lp}, and the referee for
pointing out refs.\ \cite{H,Vs}.

\end{document}